\documentclass[a4paper,12pt]{article}

\pdfoutput=1

\usepackage[tbtags]{amsmath}
\usepackage{amssymb}
\usepackage{ifthen}
\usepackage{slashed}
\usepackage{amsfonts}
\usepackage{mathrsfs}
\usepackage{bm}
\usepackage{graphicx,subfigure,booktabs}
\usepackage[numbers,sort&compress]{natbib}
\usepackage{verbatim}
\usepackage{appendix}

\usepackage[colorlinks,
            linkcolor=black,
            filecolor=black,
            anchorcolor=black,
            urlcolor=black,
            citecolor=blue
            ]{hyperref}

\usepackage{color}
\usepackage{ulem}
\usepackage{setspace}
\numberwithin{equation}{section}

\newlength{\dinwidth}
\newlength{\dinmargin}
\makeatletter
\newcommand{\thickhline}{%
    \noalign {\ifnum 0=`}\fi \hrule height 1pt
    \futurelet \reserved@a \@xhline
}
\makeatother

\allowdisplaybreaks

\begin{document}

\title{}

\title{\bf QCD analysis of electromagnetic Dalitz decays \boldmath{$J/\psi\rightarrow\eta^{(\prime)}\ell^{+}\ell^{-}$}}

\author{Jun-Kang He\; and
Chao-Jie~Fan\footnote{fancj@hbnu.edu.cn}\\[15pt]
{\small College of Physics and Electronic Science, Hubei Normal University, Huangshi 435002, China}}
\date{}


\maketitle
\vspace{0.2cm}

\begin{abstract}
{\noindent} The electromagnetic Dalitz decays $J/\psi\rightarrow\eta^{(\prime)}e^{+}e^{-}$ with large recoil momentum are studied in the framework of perturbative QCD. Meanwhile, the soft contributions from the small recoil momentum region are described by the overlap of soft wave functions, and the resonance contributions are estimated by the vector meson dominance model. Based on this dynamical picture, the transition form factors $f_{\psi\eta^{(\prime)}}(q^{2})$ in full kinematic region are calculated for the first time, and we find that the transition form factors are insensitive to the shapes of $\eta^{(\prime)}$ distribution amplitudes. Our prediction of the normalized transition form factor $F_{\psi \eta}(q^{2})\equiv f_{\psi\eta}(q^{2})/f_{\psi\eta}(0)$ agrees well with its experimental data. In addition, we also find that the branching ratios $\mathcal{B}(J/\psi\rightarrow\eta^{(\prime)}e^{+}e^{-})$ are dominated by the contributions of perturbative QCD, and the resonance contributions are negligibly small as well as the soft contributions due to the suppression of the kinematic factor. With all these contributions, our results of the branching ratios $\mathcal{B}(J/\psi\rightarrow\eta^{(\prime)}e^{+}e^{-})$ and the ratio $R_{J/\psi}^{e}=\mathcal{B}(J/\psi\rightarrow\eta e^{+}e^{-})/\mathcal{B}(J/\psi\rightarrow\eta^{\prime}e^{+}e^{-})$ are in good agreement with their experimental data. Using the obtained $F_{\psi \eta^{(\prime)}}(q^{2})$, we give the predictions of the branching ratios $\mathcal{B}(J/\psi\rightarrow\eta^{(\prime)}\mu^{+}\mu^{-})$ and their ratio $R_{J/\psi}^{\mu}$.

\end{abstract}


\newpage

\section{Introduction}
\label{sec:intro}

The decays of charmonia into light hadrons have received a great deal of attention in the past few decades both experimentally and theoretically, since they provide us with invaluable information on the strong interactions between quarks and gluons. In the quantum chromodynamics (QCD) picture, these decays are expected to proceed predominantly via $c\bar{c}$ annihilation with an intermediate state containing only gluons~\cite{Lepage:1980fj,Brodsky:1981kj,Chernyak:1983ej,Voloshin:2007dx}, so they are ideal for the study of light hadron production mechanisms, and the involved dynamical information can be extracted. In recent years, several groups have revisited the radiative decays $J/\psi \rightarrow\gamma\eta^{(\prime)}$~\cite{Ma:2002ww,Yang:2004wy,Li:2005ug,Gao:2006,Li:2007dq,He:2019mpy} as well as $h_{c} \rightarrow\gamma\eta^{(\prime)}$~\cite{Zhu:2016udl,Fan:2019sap,He:2020kin} in the framework of perturbative QCD. On the one hand, these processes are closely related to the issue of $\eta-\eta^{\prime}$ mixing, which could shed light on the $U(1)_{A}$ anomaly and the $SU(3)_{F}$ breaking. On the other hand, these processes provide a relatively clean environment to study the gluonic content of $\eta^{(\prime)}$, since there is no complication of interactions between the final light hadrons. Furthermore, these investigations show that the perturbative QCD predictions are reliable in the charmonium physics.

Recently, the BESIII Collaboration has updated the measurements of the electromagnetic (EM) Dalitz decays $J/\psi\rightarrow\eta^{(\prime)}e^{+}e^{-}$ with the branching ratios $\mathcal{B}(J/\psi\rightarrow\eta e^{+}e^{-})=(1.42\pm0.04\pm0.07)\times10^{-5}$~\cite{BESIII:2014dax,BESIII:2018qzg} and $\mathcal{B}(J/\psi\rightarrow\eta^{\prime}e^{+}e^{-})=(6.59\pm0.07\pm0.17)\times10^{-5}$~\cite{BESIII:2014dax,Ablikim:2018bhf}. As the EM Dalitz decays of light vector mesons ($\rho^{0}$, $\omega$, $\phi$), which have attracted much attention in both experiment~\cite{Achasov:2000ne,Akhmetshin:2005vy,Arnaldi:2009aa,Babusci:2014ldz,Anastasi:2016qga,Adlarson:2016hpp} and theory~\cite{Landsberg:1986fd,Reece:2009un,Branz:2009cd,Terschluesen:2010ik,Terschlusen:2011pm,Schneider:2012ez}, the $J/\psi$ decays can also be used to extract abundant information of the dynamical structure of the transition form factors (TFFs) $f_{VP}(q^{2})$ and offer a potential role in the theoretical determination of the exotic hadrons~\cite{BESIII:2021xoh} as well as the hypothetical dark photon (or, the U-boson)~\cite{Reece:2009un,Fu:2011yy,BESIII:2018qzg,Ablikim:2018bhf,Caputo:2021efm,Caputo:2021eaa}. In addition, the Dalitz decays $J/\psi\rightarrow\eta^{(\prime)}\ell^{+}\ell^{-}$ ($\ell=e,\,\mu$) are especially interesting since they involve the production of the light mesons $\eta^{(\prime)}$, which are of great phenomenological importance because of $\eta$ and $\eta^{\prime}$ mixing effects.

In the literature, the Dalitz decays $J/\psi\rightarrow\eta^{(\prime)}\ell^{+}\ell^{-}$ have been studied in different approaches~\cite{Fu:2011yy,Chen:2014yta,Zhang:2019xia}. In works~\cite{Fu:2011yy,Zhang:2019xia}, the normalized TFFs $F_{\psi \eta^{(\prime)}}(q^{2})$, which are defined as $F_{\psi \eta^{(\prime)}}(q^{2})\equiv f_{\psi \eta^{(\prime)}}(q^{2})/f_{\psi \eta^{(\prime)}}(0)$, were just parameterized as a simple pole form and the branching ratios of $J/\psi\rightarrow\eta^{(\prime)}\ell^{+}\ell^{-}$ were calculated with the vector meson dominance (VMD) model. Although the predictions of the branching ratios $\mathcal{B}(J/\psi\rightarrow\eta^{(\prime)} e^{+}e^{-})$ are compatible with the experimental measurements~\cite{BESIII:2018qzg,Ablikim:2018bhf}, their normalized TFFs $F_{\psi \eta^{(\prime)}}(q^{2})$ with the point-like particle assumption would gloss over much dynamical information from the QCD processes. While the dynamical information could offer us an insight into the nature of the Okubo-Zweig-Iizuka (OZI) rule~\cite{Okubo:1963fa,Iizuka:1966fk} and the $U_{A}(1)$ anomaly as well as the $\eta-\eta^{\prime}$ mixing~\cite{Novikov:1979uy,Chao:1989pi,Chao:1990im,Feldmann:1998vh,Yang:2004wy,He:2019mpy}. Therefore, a deep theoretical study of the TFFs $f_{\psi \eta^{(\prime)}}(q^{2})$ should consider the information of the internal structure of mesons, as well as the transition mechanisms in different kinematic regions. We will present detailed discussions in the later part of this paper. Besides the simple pole approximation, Chen {\it et al.}~\cite{Chen:2014yta} studied these Dalitz decay processes with the effective Lagrangian approach, and they confirmed that the $J/\psi\rightarrow\eta^{(\prime)}\gamma^{(\ast)}$ processes were predominantly dominated by the $J/\psi\rightarrow\eta_{c}\gamma^{(\ast)}\rightarrow\eta^{(\prime)}\gamma^{(\ast)}$ mechanism. Perhaps this may need more discussion~\cite{Feldmann:1998vh,He:2019mpy,Fan:2019sap,He:2020kin}.

In general, there are several types of contributions to the TFFs involved in the EM Dalitz decay processes $J/\psi\rightarrow\eta^{(\prime)}\ell^{+}\ell^{-}$: (i) In the large recoil momentum region, $q^{2}\simeq 0$, the contributions are dominated by the hard mechanism, which can be calculated in the framework of perturbative QCD. Just as our recent investigations~\cite{He:2019mpy,Fan:2019sap,He:2020kin}, the perturbative QCD approach can be reliably employed in the charmonia radiative decay processes $J/\psi (h_{c}) \rightarrow\gamma\eta^{(\prime)}$. (ii) In the small recoil momentum region, $q^{2}\simeq q^{2}_{\text{max}}=(M_{J/\psi}-m_{\eta^{(\prime)}})^{2}$, phenomenologically, the TFFs can be interpreted as the wave function overlap~\cite{Radyushkin:1998rt,Radyushkin:1998vb,Feldmann:1999sm,Huang:2000kd,Chang:2001pm,Zhao:2010mm}. Namely, the corresponding contributions are governed by the overlapping integration of the soft wave functions. (iii) In resonance regions, such as $q^{2}\simeq m_{\rho}^{2}, \, m_{\omega}^{2}, \, m_{\phi}^{2}$, the resonance interaction between photons and hadrons is predominant, which can be universally described by a vector meson dominance (VMD) model~\cite{Landsberg:1986fd}. In this work, the aforementioned contributions are all taken into account. In the large recoil momentum region, we adopt light-cone distribution amplitudes (DAs) to describe the internal dynamics of $\eta^{(\prime)}$ where both the quark-antiquark content and the gluonic content are taken into account, and the detailed structure of $J/\psi$ is described by its Bethe-Salpeter (B-S) wave function. We evaluate analytically the involved one-loop integrals, and find the TFFs barely depend on the light quark masses and the shapes of the light meson DAs, which is compatible with the situation in the decay processes $J/\psi \rightarrow\gamma\eta^{(\prime)}$~\cite{He:2019mpy} as well as $h_{c} \rightarrow\gamma\eta^{(\prime)}$~\cite{Fan:2019sap,He:2020kin}. In the whole kinematic region, we present a QCD analysis of the TFFs $f_{\psi\eta^{(\prime)}}(q^{2})$ for the first time, and our prediction of the normalized TFF $F_{\psi \eta}(q^{2})$ is in good agreement with the experimental data. In addition, although the VMD contributions and the soft contributions are small in the branching ratios $\mathcal{B}(J/\psi\rightarrow\eta^{(\prime)}e^{+}e^{-})$ because of a suppression of the kinematic factor, they play a significant role in the TFFs. By using the normalized TFFs $F_{\psi \eta^{(\prime)}}(q^{2})$ extracted from the decay processes $J/\psi\rightarrow\eta^{(\prime)}e^{+}e^{-}$, we obtain the predictions of the branching ratios $\mathcal{B}(J/\psi\rightarrow\eta^{(\prime)}\mu^{+}\mu^{-})$ and their ratio $R_{J/\psi}^{\mu}=\mathcal{B}(J/\psi\rightarrow\eta \mu^{+}\mu^{-})/\mathcal{B}(J/\psi\rightarrow\eta^{\prime}\mu^{+}\mu^{-})$.

The paper is organized as follows. The theoretical framework for the decay processes $J/\psi\rightarrow\eta^{(\prime)}\ell^{+}\ell^{-}$ is shown in detail in section~\ref{sec:framework}. In section~\ref{sec:Results and discussions} we present our numerical results and some phenomenological discussions, and the last section is our summary.

\section{Theoretical framework}
\label{sec:framework}
\subsection{Hard mechanism}
\label{subsec:Hard mechanism}
\subsubsection{The contributions of the quark-antiquark content of $\eta^{(\prime)}$}
\label{subsubsec:QCDq}

\begin{figure}[th]
  \begin{center}
  \includegraphics[width=0.4\textwidth]{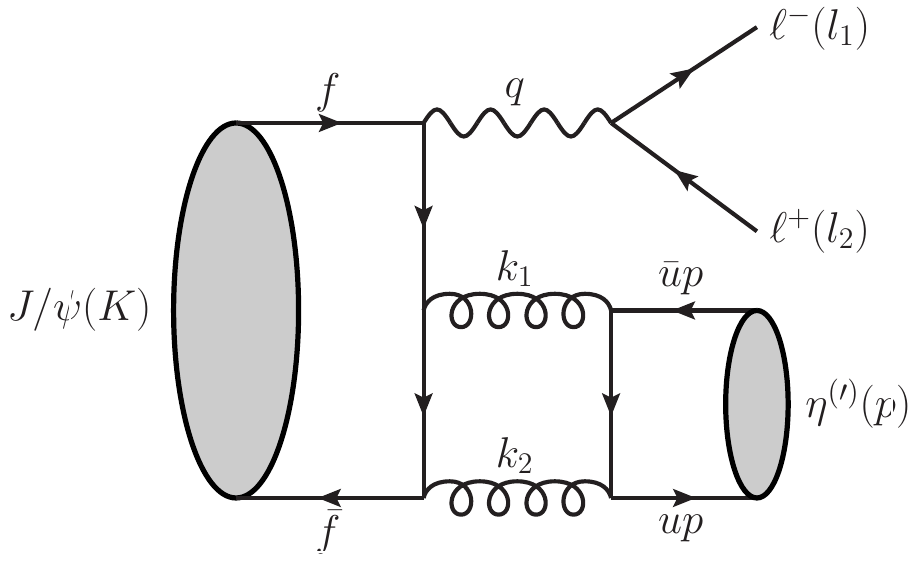}
  \end{center}
  \vskip -0.7cm
  \caption{One typical Feynman diagram for $J/\psi\rightarrow\eta^{(\prime)}\ell^{+}\ell^{-}$ with the quark-antiquark content of $\eta^{(\prime)}$. The kinematical variables are labeled.}\label{QCDq}
\end{figure}

For the quark-antiquark content of $\eta^{(\prime)}$, one of the leading order
Feynman diagrams for the decay processes $J/\psi\rightarrow\eta^{(\prime)}\ell^{+}\ell^{-}$ is depicted in Fig.~\ref{QCDq}, and the other five diagrams arise from permutations of the photon and the gluon legs. Here $f$ and $\bar{f}$ represent the momenta of the charm quark and the charm antiquark respectively, $k_{1}$ and $k_{2}$ represent the momenta of gluons, $p$ represents the momentum of $\eta^{(\prime)}$, $u$ and $\bar{u}$ are the momentum fractions carried by the light quark and the light antiquark respectively. According to the Feynman diagrams, we can easily give the amplitude of $J/\psi\rightarrow\eta^{(\prime)}\ell^{+}\ell^{-}$:
\begin{eqnarray}
{\mathcal M}=-\frac{e}{q^{2}}{\mathcal A}^{\alpha\beta}\varepsilon_{\alpha}(K)\bar{u}(l_{1})\gamma_{\beta} v(l_{2}),
\end{eqnarray}
where $\mathcal{A}^{\alpha\beta}$ represents the amplitude of $J/\psi\rightarrow\eta^{(\prime)}\gamma^{\ast}$, $K$ and $\varepsilon(K)$ are the momentum and polarization vector of $J/\psi$ respectively, $q$ is the momentum of the virtual photon, $q^{2}=m_{\ell^{+}\ell^{-}}^{2}$ is the square of the invariant mass of the lepton pair, $l_{1}$ and $l_{2}$ are the momenta of the leptons $\ell^{-}$ and $\ell^{+}$ respectively.

Following the method developed in Refs.~\cite{Kuhn:1979bb,Guberina:1980xb,Korner:1982vg}, we divide the amplitude of $J/\psi\rightarrow \eta^{(\prime)}\gamma^{\ast}$ into two parts. One part describes the effective coupling between $J/\psi$, a virtual photon and two virtual gluons, i.e., the process $J/\psi\rightarrow g^{\ast}g^{\ast} \gamma^{\ast}$. The other part describes the effective coupling between $\eta^{(\prime)}$ and two virtual gluons, i.e., the processes $g^{\ast}g^{\ast}\rightarrow\eta^{(\prime)}$. Then one just multiplies the two parts, inserts the gluon propagators and performs the loop integrations to obtain the final amplitude of $J/\psi\rightarrow \eta^{(\prime)}\gamma^{\ast}$.

In the rest frame of $J/\psi$, one can write the amplitude of $J/\psi\rightarrow g^{\ast}g^{\ast} \gamma^{\ast}$ in the form~\cite{Guberina:1980dc,Guberina:1980xb,Korner:1982vg}
\begin{eqnarray}\label{ggr}
{\mathcal A}^{\alpha\beta\mu\nu}_{1}\varepsilon_{\alpha}(K)\epsilon^{*}_{\beta}(q)\epsilon^{*}_{\mu}(k_{1})\epsilon^{*}_{\nu}(k_{2})
&=&\sqrt{3}\int\frac{\mathrm{d}^{4}k}{(2\pi)^{4}}\textrm{Tr}\left[\chi(K,k){\cal O}(k)\right],
\end{eqnarray}
where $\chi(K, k)$ is the B-S wave function of $J/\psi$ and $\mathcal{O}(k)$ is the hard-scattering amplitude. Here $\sqrt{3}$ is the color factor and $\epsilon(q)$ is the polarization vector of the virtual photon. $k_{1}$, $k_{2}$ and $\epsilon(k_{1})$, $\epsilon(k_{2})$ are the momenta and polarization vectors of the two gluons, respectively. The momenta of the quark $c$ and antiquark $\bar{c}$ read
\begin{eqnarray}
f=\frac{K}{2}+k,\quad\quad   \bar{f}=\frac{K}{2}-k
\end{eqnarray}
with $k$ the relative momentum between the quark $c$ and antiquark $\bar{c}$. In a nonrelativistic bound state picture, one can reduce the B-S wave function $\chi(K, k)$ to its nonrelativistic form
~\cite{Kuhn:1979bb,Guberina:1980dc}
\begin{eqnarray}\label{bsfun}
\chi(K, k)=2\pi\delta(k^{0})\psi_{00}(\boldsymbol {k})\left[\sqrt{\frac{1}{4M}}\slashed{\varepsilon}(K)(\slashed{K}-M)\right],
\end{eqnarray}
where $\psi_{00}(\boldsymbol {k})$ is the bound state wave function of $S$-wave charmonium $J/\psi$, and $M$ is the mass of $J/\psi$. For $S$-wave charmonium decays, one could neglect the dependence of the hard-scattering amplitude $\mathcal{O}(k)$ on the relative momentum $k$ in the leading order approximation~\cite{Barbieri:1979iy,Guberina:1980dc,He:2019mpy}:
\begin{eqnarray}\label{hardo}
\mathcal{O}(k)\simeq \mathcal{O}(0).
\end{eqnarray}
And the higher order corrections related to the relative momentum are negligible, because the B-S wave function of charmonium is heavily damped on the relative momentum. Using the Fourier transformation of the bound state wave function, one obtains the well-known result in coordinate space
\begin{eqnarray}\label{Fouriertr}
\int \frac{\mathrm{d}^{3}k}{(2\pi)^{3}} \psi_{00}(\boldsymbol {k})=\sqrt{\frac{1}{4\pi}} R_{\psi}(0).
\end{eqnarray}
With the help of the B-S wave function Eq.~\eqref{bsfun} and the hard-scattering amplitude Eq.~\eqref{hardo} as well as the Fourier transformation Eq.~\eqref{Fouriertr}, the amplitude of $J/\psi\rightarrow g^{\ast}g^{\ast}\gamma^{\ast}$ can be rewritten as
\begin{eqnarray}
{\mathcal A}^{\alpha\beta\mu\nu}_{1}\varepsilon_{\alpha}(K)\epsilon^{*}_{\beta}(q)\epsilon^{*}_{\mu}(k_{1})\epsilon^{*}_{\nu}(k_{2})
=\frac{1}{2}\sqrt{\frac{3}{4\pi M}}R_{\psi}(0)\textrm{Tr}\left[\slashed{\varepsilon}(K)(\slashed{K}-M){\cal O}(0)\right],
\end{eqnarray}
where the hard-scattering amplitude ${\cal O}(0)$ reads
\begin{eqnarray}
\mathcal{O}(0) &=& iQ_{c}eg_{s}^{2}\frac{\delta_{ab}}{6}\slashed{\epsilon}^{\ast}(k_{2}) \frac{\slashed{k}_{2}-\slashed{q}
-\slashed{k}_{1}+M}{-2 (q+k_{1})\cdot k_{2}}\slashed{\epsilon}^{\ast}(q)\frac{\slashed{k}_{2}+\slashed{q}-\slashed{k}_{1}+M}
{-2 (q+k_{2})\cdot k_{1}} \slashed{\epsilon}^{\ast}(k_{1}) \nonumber\\
& &+\textrm{(5~permutations~of~ $k_{1}$,~$k_{2}$~and~$q$)}.
\end{eqnarray}

In what follows, we make a brief summary of the coupling $g^{\ast}g^{\ast}-\eta^{(\prime)}$. At the leading twist level, the light-cone DA of $\eta^{(\prime)}$ is defined according to~\cite{Chernyak:1983ej,Ali:2003kg,Ball:2007hb}
\begin{eqnarray}\label{matri}
\langle \eta^{(\prime)}(p) |\bar{q}_{\alpha}(x)q_{\beta}(y)|0\rangle &=&\frac{i}{4}f_{\eta^{(\prime)}}^{q}\left(\slashed{p}\gamma_{5}\right)_{\beta\alpha}
 \int\textrm{d}u e^{i(\bar{u}p\cdot y+up\cdot x)}\phi^{q}(u),
\end{eqnarray}
where the superscript $q$ denotes the light quark ($q=u,d,s$). The decay constants $f_{\eta^{(\prime)}}^{q}$ are defined according to
\begin{eqnarray}
\langle0|\bar{q}(0)\gamma_{\mu}\gamma_{5}q(0)|\eta^{(\prime)}(p)\rangle&=&
i f_{\eta^{(\prime)}}^{q}p_{\mu}.
\end{eqnarray}
Using Eq.~\eqref{matri}, we obtain the amplitude of $g^{\ast}g^{\ast}\rightarrow\eta^{(\prime)}$~\cite{Muta:1999tc,Yang:2000ce,Ali:2000ci}:
\begin{eqnarray}
\mathcal{A}^{\mu\nu}_{2}&=&-i (4\pi \alpha_{s})\delta_{ab}\epsilon^{\mu\nu\rho\sigma}k_{1\rho}k_{2\sigma}\nonumber\\
& &\sum_{q=u,d,s}\frac{f_{\eta^{(\prime)}}^{q}}{6}
\int^{1}_{0}du\phi^{q}(u)\left(\frac{1}{\bar{u}k_{1}^{2}+uk_{2}^{2}-u\bar{u}m^{2}-m_{q}^{2}}+(u\leftrightarrow\bar{u})\right).
\end{eqnarray}
Here $\bar{u}=1-u$, $u$ is the momentum fraction carried by the quark, $m_{q}$ is the mass of the quark ($q=u,d,s$), $m$ is the mass of $\eta^{(\prime)}$.  The light-cone DA is~\cite{Agaev:2014wna}
\begin{eqnarray}
\phi^{q}(u)&=&6u(1-u)\left[1+\sum_{n=2,4\cdots}c^{q}_{n}(\mu)C_{n}^{\frac{3}{2}}(2u-1)\right]
\end{eqnarray}
with $c^{q}_{n}(\mu)$ the Gegenbauer moments, and we take its three models listed in Table 1 of Refs.~\cite{He:2019mpy,Fan:2019sap}. We find that the TFFs barely depend on the shapes of $\eta^{(\prime)}$ DAs (we will estimate them below).

Then the decay amplitude of $J/\psi\rightarrow \eta^{(\prime)}\gamma^{\ast}$ can be obtained by contracting the above two couplings, inserting the gluon propagators and integrating over the loop momentum (see~\cite{Guberina:1980xb,Korner:1982vg} for more details)
\begin{eqnarray}
\mathcal{A}^{\alpha\beta}\varepsilon_{\alpha}(K)\epsilon^{\ast}_{\beta}(q)=\frac{1}{2}\int\frac{\mathrm{d}^{4}k_{1}}{(2\pi)^{4}}{\mathcal A}^{\alpha\beta\mu\nu}_{1}\mathcal{A}_{2\mu\nu}\frac{i}{k^{2}_{1}
+i\epsilon}\frac{i}{k^{2}_{2}+i\epsilon}\varepsilon_{\alpha}(K)\epsilon^{\ast}_{\beta}(q),
\end{eqnarray}
i.e.,
\begin{eqnarray}\label{aalph}
\mathcal{A}^{\alpha\beta}=\frac{1}{2}\int\frac{\mathrm{d}^{4}k_{1}}{(2\pi)^{4}}{\mathcal A}^{\alpha\beta\mu\nu}_{1}\mathcal{A}_{2\mu\nu}\frac{i}{k^{2}_{1}
+i\epsilon}\frac{i}{k^{2}_{2}+i\epsilon}.
\end{eqnarray}
Considering parity conservation, Lorentz invariance and gauge invariance, one knows
\begin{eqnarray}
\mathcal{A}^{\alpha\beta}\propto \left(\epsilon^{\alpha\beta\mu\nu}p_{\mu}q_{\nu}\right).
\end{eqnarray}
Then the $J/\psi\rightarrow \eta^{(\prime)}\gamma^{\ast}$ TFFs are defined by
\begin{eqnarray}
\mathcal{A}^{\alpha\beta}=-e f^{Q}_{\psi \eta^{(\prime)}}(q^{2}) \epsilon^{\alpha\beta\mu\nu}p_{\mu}q_{\nu}.
\end{eqnarray}
With the help of the projection operator
\begin{eqnarray}
\mathcal{P}^{\alpha\beta}=\frac{\epsilon^{\alpha\beta\mu\nu}p_{\mu}q_{\nu}}{\lambda^{\frac{1}{2}}(M^{2},\,m^{2},\,q^{2})}
\end{eqnarray}
and the normalization condition
\begin{eqnarray}
\mathcal{P}^{\alpha\beta}\mathcal{P}_{\alpha\beta}=\frac{1}{2},
\end{eqnarray}
the TFFs can be rewritten as
\begin{eqnarray}\label{fpsiet}
f^{Q}_{\psi \eta^{(\prime)}}(q^{2})=-\frac{2e^{-1}}{\lambda^{\frac{1}{2}}(M^{2},\,m^{2},\,q^{2})}\mathcal{P}_{\alpha\beta}\mathcal{A}^{\alpha\beta}
\end{eqnarray}
with $\lambda(a,\,b,\,c)\equiv a^{2}+b^{2}+c^{2}-2(ab+bc+ac)$ the usual K\"{a}ll\'{e}n function. Here we show the expression of the TFFs more clearly
\begin{eqnarray}\label{analyfpsiet}
f^{Q}_{\psi \eta^{(\prime)}}(q^{2})&=&\frac{16 R_{\psi}(0)}{\lambda(M^{2},\,m^{2},\,q^{2})}\frac{Q_{c}(4\pi \alpha_{s})^{2}}{3\sqrt{3}}\sqrt{\frac{M}{\pi}}\sum_{q}f_{\eta^{(\prime)}}^{q}\int\mathrm{d}u \phi^{q}(u)\int\frac{\mathrm{d}^{4}k_{1}}{(2\pi)^{4}}\frac{ \left(k_{1}^{2}-k_{1}\cdot p\right)}{\left(M^{2}-m^{2}-q^{2}\right)}\nonumber\\
&&\times\bigg[\Big(\frac{k_{1}\cdot p \left(M^{2}-m^{2}-q^{2}\right) \left(4 k_{1}\cdot q+M^{2}-m^{2}-q^{2}\right)}{2 D_{1}D_{2}D_{3}D_{4}D_{5}}-\frac{8 q^{2} k_{1}\cdot p^{2}}{2 D_{1}D_{2}D_{3}D_{4}D_{5}}\nonumber\\
&&-\frac{2 m^{2} k_{1}\cdot q \left(M^{2}-m^{2}-q^{2}\right)}{2 D_{1}D_{2}D_{3}D_{4}D_{5}}-\frac{\lambda(M^{2},\,m^{2},\,q^{2})}{D_{2}D_{3}D_{4}D_{5}}\Big)+(u \leftrightarrow \bar{u})\bigg],
\end{eqnarray}
where the expressions of the denominators read
\begin{eqnarray}
D_{1}&=&k_{1}^{2}+i \epsilon\nonumber\\
D_{2}&=&(k_{1}-p)^{2}+i \epsilon\nonumber\\
D_{3}&=&(k_{1}-u p)^{2}-m_{q}^{2}+i \epsilon\nonumber\\
D_{4}&=&\frac{1}{4}\left[(2 k_{1}-p-q)^{2}-M^{2}\right]+i \epsilon\nonumber\\
D_{5}&=&\frac{1}{4}\left[(2 k_{1}-p+q)^{2}-M^{2}\right]+i \epsilon.
\end{eqnarray}
By using the algebraic identity
\begin{eqnarray}
1=\frac{2(D_{1}+D_{2}-D_{4}-D_{5})}{M^{2}+m^{2}-q^{2}},
\end{eqnarray}
the TFFs $f^{Q}_{\psi \eta^{(\prime)}}(q^{2})$ in Eq.~\eqref{analyfpsiet} can be decomposed into sum of four-point one-loop integrals, and then it can be analytically evaluated with the technique proposed in Refs.~\cite{tHooft:1978jhc,Denner:1991qq,Denner:1991kt} or the computer program $Package-\mathrm{X}$~\cite{Patel:2015tea,Patel:2016fam}. By integrating over the loop momentum $k_{1}$ and the momentum fraction $u$, we find that the TFFs $f^{Q}_{\psi \eta^{(\prime)}}(q^{2})$ are very insensitive to the light quark mass $m_{q}$ as well as the shapes of $\eta^{(\prime)}$ DAs. This is similar to the situations of the dimensionless functions involved in the radiative decays $J/\psi (h_{c}) \rightarrow\gamma\eta^{(\prime)}$~\cite{He:2019mpy,Fan:2019sap,He:2020kin}. Specifically, the change of the modulus of the TFFs $f^{Q}_{\psi \eta^{(\prime)}}(q^{2})$ does not exceed $1\%$ when the value of the light quark mass $m_{q}$ varies in the range $(0-100)\, \mathrm{MeV}$ with the different models of the DAs. Therefore, the theoretical uncertainties from the DAs are ignorable in our calculations of the TFFs.

\begin{figure}[th]
  \begin{center}
  \includegraphics[width=0.4\textwidth]{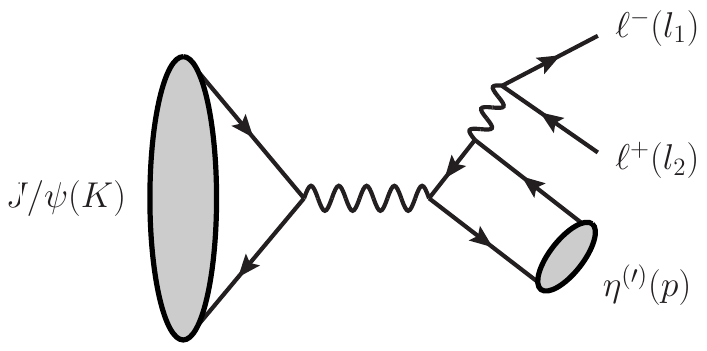}~~~~~~
  \includegraphics[width=0.4\textwidth]{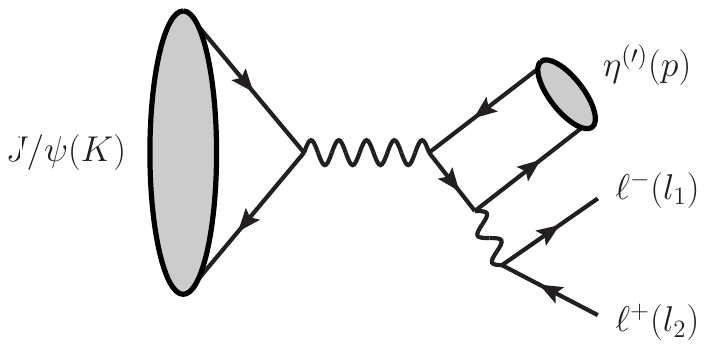}
  \end{center}
  \vskip -0.7cm
  \caption{Feynman diagrams for the QED processes $J/\psi\rightarrow\gamma^{\ast}\rightarrow\eta^{(\prime)}\ell^{+}\ell^{-}$. The kinematical variables are labeled.}\label{QED}
\end{figure}

It is worthwhile to point out that the QED processes $J/\psi\rightarrow\gamma^{\ast}\rightarrow\eta^{(\prime)}\ell^{+}\ell^{-}$ can also contribute to the EM Dalitz decays $J/\psi\rightarrow\eta^{(\prime)}\ell^{+}\ell^{-}$ and the corresponding Feynman diagrams are shown in Fig.~\ref{QED}.
Then the corresponding TFFs $f^{E}_{\psi \eta^{(\prime)}}(q^{2})$ can be expressed as
\begin{eqnarray}
f^{E}_{\psi \eta^{(\prime)}}(q^{2})&=&i \sqrt{3} Q_{c}(4\pi \alpha)\frac{R_{\psi}(0)}{M^{2}}\sqrt{\frac{M}{\pi}} \sum_{q=u,\,d,\,s}Q_{q}^{2}f_{\eta^{(\prime)}}^{q}\int\mathrm{d}u \phi^{q}(u)\nonumber\\
&&\times\left[\frac{1}{q^{2}+u^{2}m^{2}+u(M^{2}-m^{2}-q^{2})-m_{q}^{2}+i\epsilon}+(u\leftrightarrow\bar{u})\right],
\end{eqnarray}
where the $Q_{q}$ represents the light quark charge.

\subsubsection{The contributions of the gluonic content of $\eta^{(\prime)}$}
\label{subsubsec:QCDg}
\begin{figure}[th]
  \begin{center}
  \includegraphics[width=0.4\textwidth]{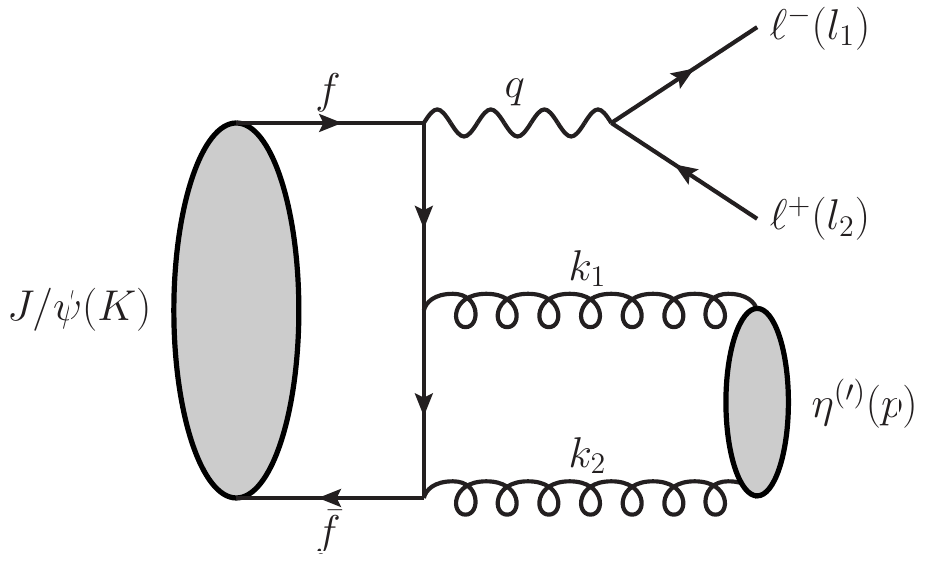}
  \end{center}
  \vskip -0.7cm
  \caption{One typical Feynman diagram for $J/\psi\rightarrow\eta^{(\prime)}\ell^{+}\ell^{-}$ with the gluonic content of $\eta^{(\prime)}$. The kinematical variables are labeled.}\label{QCDg}
\end{figure}

As we have emphasized in Ref.~\cite{He:2019mpy}, although the contributions of the gluonic content of $\eta^{(\prime)}$ in the radiative decay processes $J/\psi\rightarrow\eta^{(\prime)}\gamma$ can directly come from the tree level, the amplitudes are strongly suppressed by the factor $m^{2}/M^{2}$. As a consequence, the gluonic contributions only offer small corrections. Obviously, the situation should be found in the Dalitz decay processes $J/\psi\rightarrow\eta^{(\prime)}\ell^{+}\ell^{-}$, because the Dalitz decay processes and the corresponding radiative decay processes have the same spin structures in their hadronic matrix elements. The typical Feynman diagram is exhibited in Fig.~\ref{QCDg}, and there are the other two diagrams from permutations of the photon and the gluon legs.

At the leading twist level, the light-cone matrix elements of the meson $\eta^{(\prime)}$ over two-gluon fields can be written as~\cite{Kroll:2002nt,Ball:2007hb,Agaev:2014wna}:
\begin{eqnarray}
\langle\eta^{(\prime)}(p)|A_{\alpha}^{a}(x)A_{\beta}^{b}(y)|0\rangle=\frac{1}{4}\epsilon_{\alpha\beta\mu\nu}
\frac{n^{\mu}p^{\nu}}{p\cdot n}\frac{C_{F}}{\sqrt{3}}\frac{\delta^{ab}}{8}f_{\eta^{(\prime)}}^{1}\int\textrm{d}u e^{i(up\cdot x+\bar{u}p\cdot y)}\frac{\phi^{g}(u)}{u(1-u)},
\end{eqnarray}
where $n=(0,\,1,\,\mathbf{0}_{\perp})$ is a lightlike vector~\cite{Kroll:2002nt}. Here $f_{\eta^{(\prime)}}^{1}=\frac{1}{\sqrt{3}}(f_{\eta^{(\prime)}}^{u}+f_{\eta^{(\prime)}}^{d}+f_{\eta^{(\prime)}}^{s})$ are the effective decay constant and the gluonic twist-2 DA reads~\cite{Agaev:2014wna,Ball:2007hb,Alte:2015dpo}
\begin{eqnarray}
\phi^{g}(u)=30u^{2}(1-u)^{2}\sum_{n=2,4\cdots}c^{g}_{n}(\mu)C_{n-1}^{\frac{5}{2}}(2u-1).
\end{eqnarray}

After a series of calculations, the corresponding TFFs $f^{G}_{\psi \eta^{(\prime)}}(q^{2})$ can be expressed as
\begin{eqnarray}\label{fg}
f^{G}_{\psi \eta^{(\prime)}}(q^{2}) &=& i \frac{8 R_{\psi}(0)}{\lambda(M^{2},\,m^{2},\,q^{2})}\frac{Q_{c}(4\pi \alpha_{s})}{9}\sqrt{\frac{M}{\pi}}\nonumber\\
&&\times f_{\eta^{(\prime)}}^{1}\int\mathrm{d}u \frac{\phi^{g}(u)}{u(1-u)}
\frac{m^{2}(M^{2}-m^{2}-q^{2})(1-2u)}{\left[(M^{2}-q^{2})^{2}-m^{4}(1-2u)^{2}\right]}.
\end{eqnarray}
Clearly, The TFFs are suppressed by $m^{2}$ (namely, a dimensionless factor $m^{2}/M^{2}$). As pointed out in Ref.~\cite{He:2019mpy}, the leading twist gluonic content contributions mainly come from two on-shell gluons, which give a suppression factor $m^{2}/M^{2}$ due to the special form of the Ore-Powell matrix elements~\cite{Krammer:1978qp,Billoire:1978xt}. From the point of view of the QCD evolution of the gluon DA~\cite{Ali:2000ci,Kroll:2002nt,Ali:2003kg}, the contributions from the gluonic content of $\eta^{(\prime)}$ are supposed to be small since the gluonic content can be seen as the higher order effects.

Based on the foregoing discussions, in the large recoil momentum region, the $J/\psi\rightarrow \eta^{(\prime)}\gamma^{\ast}$ TFFs can be obtained by
\begin{eqnarray}
f^{H}_{\psi \eta^{(\prime)}}(q^{2})=f^{Q}_{\psi \eta^{(\prime)}}(q^{2})+f^{E}_{\psi \eta^{(\prime)}}(q^{2})+f^{G}_{\psi \eta^{(\prime)}}(q^{2}),
\end{eqnarray}
which include the dynamical structure information from the QCD and the QED processes. And all these contributions can be reliably calculated in the framework of the hard mechanism.

\subsection{Soft mechanism}
\label{subsec:Soft mechanism}
In the small recoil momentum region, the perturbative QCD approach becomes invalid in the EM Dalitz decay processes $J/\psi\rightarrow\eta^{(\prime)}\ell^{+}\ell^{-}$. And the dominant contributions to the TFFs are controlled by the soft mechanism, which can be treated as the soft wave function overlap. Phenomenologically, one can adopt an empirical form factor~\cite{Close:2000yk,Zhao:2010mm}:
\begin{eqnarray}
f^{S}_{\psi \eta^{(\prime)}}(q^{2})=g_{\psi \eta^{(\prime)}}\exp\left(-\frac{\boldsymbol{q}^{2}}{8\beta^{2}}\right),
\end{eqnarray}
where $g_{\psi \eta^{(\prime)}}$ denote the $J/\psi-\eta^{(\prime)}-\gamma^{\ast}$ coupling and are determined by the continuity condition of the TFFs between the large and the small recoil momentum regions, and the parameter $\beta$ is in a range of $(300-500)\, \mathrm{MeV}$~\cite{Close:2000yk,Zhao:2010mm}. Our numerical analysis of the TFF $f_{\psi \eta}(q^{2})$ indicates that a smaller value $\beta=370\, \mathrm{MeV}$ is favored, which may be due to that the intermediate photon is in a highly virtual kinematic region and part of the off-shell effects would be absorbed into the parameter $\beta$. The smaller value is compatible with the result in Ref~\cite{Zhao:2010mm}. In addition, the branching ratios $\mathcal{B}(J/\psi\rightarrow\eta^{(\prime)}\ell^{+}\ell^{-})$ are insensitive to the parameter $\beta$ due to a strong suppression of the kinematic factor in the small recoil momentum region.

In the whole recoil momentum region, the TFFs can be expressed as
\begin{eqnarray}\label{TFFs}
f_{\psi \eta^{(\prime)}}(q^{2})=
\begin{cases}
f^{H}_{\psi \eta^{(\prime)}}(q^{2})      ~~~~& q^{2}\leq 1\, \mathrm{GeV}^{2} , \\
f^{S}_{\psi \eta^{(\prime)}}(q^{2})      ~~~~& q^{2}>1\, \mathrm{GeV}^{2} .
\end{cases}
\end{eqnarray}
It is worth reminding that the recoil momentum of $\eta^{(\prime)}$ is above $1\, \mathrm{GeV}$ when $q^{2}\leq 1\, \mathrm{GeV}^{2}$, and below or near $1\, \mathrm{GeV}$ when $q^{2}> 1\, \mathrm{GeV}^{2}$. Generally speaking~\cite{Radyushkin:1990te,Jakob:1993iw,Jakob:1994hd,Bolz:1997ez}, perturbative QCD begins to be self-consistent at the recoil momentum as low as $1\, \mathrm{GeV}$, i.e., the transition to perturbative QCD appears at about $q^{2}= 1\, \mathrm{GeV}^{2}$, where the hard mechanism begins to dominate as the $q^{2}$ decreases. Although we clearly separate the hard contributions in the large recoil momentum region and the soft ones in the small recoil momentum region, how to precisely match these two contributions in the intermediate recoil momentum region still needs further investigations. Even so, our description of the decay processes $J/\psi\rightarrow \eta^{(\prime)}\gamma^{\ast}$ may constitute an important step forward towards a satisfactory description.

\subsection{VMD model}
\label{subsec:VMD model}
\begin{figure}[th]
  \begin{center}
  \includegraphics[width=0.4\textwidth]{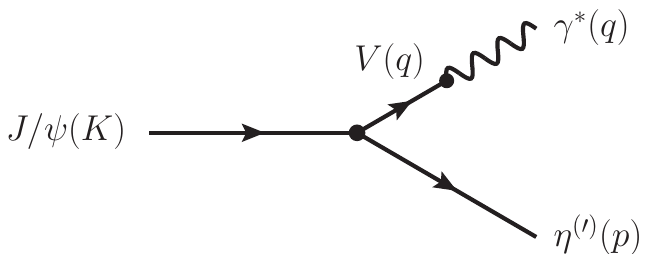}
  \end{center}
  \vskip -0.7cm
  \caption{Schematic diagram for $J/\psi\rightarrow\eta^{(\prime)}\gamma^{\ast}$ in the framework of VMD. The kinematical variables are labeled.}\label{VMD}
\end{figure}

In this subsection we will briefly discuss the resonance interaction, which can be described by VMD model. The VMD contributing diagram is illustrated in Fig.~\ref{VMD}, where $V$ represents vector mesons. Here we concentrate on the resonance that is closest to the physical decay region, in which it mainly includes the light mesons $\rho$, $\omega$ and $\phi$.

The effective Lagrangian for $J/\psi-V-P$ coupling can be written as~\cite{Intemann:1983yj,Zhao:2006gw,Zhao:2010mm,Lichard:2010ap}:
\begin{eqnarray}\label{eLagrangian}
\mathcal{L}_{\psi VP}=\frac{g_{\psi VP}(q^{2})}{M}\epsilon_{\alpha\beta\mu\nu}\partial^{\alpha}\psi^{\beta}\partial^{\mu}V^{\nu}P,
\end{eqnarray}
where $V$ ($V=\rho,\,\omega,\,\phi$), $\psi$ ($\psi=J/\psi$) and $P$ ($P=\eta^{(\prime)}$) are the corresponding vector and pseudoscalar meson fields, $g_{\psi VP}(q^{2})=g_{\psi VP}\exp\left(\frac{-\boldsymbol{q}^{2}}{8\beta^{2}}\right)$ is dimensionless coupling parameter (see Refs.~\cite{Zhao:2006gw,Zhao:2010mm} for more details) and the undetermined constant $g_{\psi VP}$ can be determined by the decay process $J/\psi\rightarrow VP$. Following the effective Lagrangian of Eq~\eqref{eLagrangian}, one can easily derive the undetermined constant:
\begin{eqnarray}\label{gpsiVP}
g_{\psi VP} =\left(\frac{96\pi M^{5}\Gamma^{exp}_{J/\psi\rightarrow VP}}
{\lambda^{\frac{3}{2}}(M^{2},\,m_{V}^{2},\,m_{P}^{2})}\right)^{\frac{1}{2}}
\exp\left(\frac{\lambda(M^{2},\,m_{V}^{2},\,m_{P}^{2})}{32M^{2}\beta^{2}}\right),
\end{eqnarray}
where $m_{V}$ is the mass of the vector meson $V$, $m_{P}$ is the mass of the pseudoscalar meson $P$.

The effective Lagrangian for $V-\gamma^{\ast}$ coupling can be described as~\cite{Bauer:1977iq,Zhao:2006gw,Zhao:2010mm}:
\begin{eqnarray}
\mathcal{L}_{V\gamma^{\ast}}=\frac{e m_{V}^{2}}{f_{V}}V_{\mu}A^{\mu},
\end{eqnarray}
where $e m_{V}^{2}/f_{V}$ is the photon-vector-meson coupling constant, $A$ denotes the EM field. The undetermined constant $f_{V}$ can be extracted from the decay process $V\rightarrow e^{+}e^{-}$:
\begin{eqnarray}
\mid f_{V} \mid=\left(\frac{4\pi \alpha^{2}m_{V}}{3\Gamma^{exp}_{V\rightarrow e^{+}e^{-}}}\right)^{\frac{1}{2}}.
\end{eqnarray}

Then the corresponding TFFs can be read as
\begin{eqnarray}\label{TFFVMD}
f^{V}_{\psi \eta^{(\prime)}}(q^{2})=-i\frac{g_{\psi V\eta^{(\prime)}}m_{V}^{2}}{Mf_{V}(q^{2}-m_{V}^{2}+im_{V}\Gamma_{V})}\exp\left(\frac{-\boldsymbol{q}^{2}}{8\beta^{2}}\right),
\end{eqnarray}
where $\Gamma_{V}$ is the full width of the vector meson $V$. When the intermediate vector meson is near the on-mass-shell, we obtain
\begin{eqnarray}
f^{V}_{\psi \eta^{(\prime)}}\sim \left(\frac{\Gamma^{exp}_{J/\psi\rightarrow V\eta^{(\prime)}}}{\Gamma_{V}}\right)^{\frac{1}{2}}\mathcal{B}^{exp}(V\rightarrow e^{+}e^{-})^{\frac{1}{2}}.
\end{eqnarray}
It is worth noting that the TFFs $f^{\rho}_{\psi \eta^{(\prime)}}$ are an order of magnitude smaller than the TFFs $f^{\omega,\,\phi}_{\psi \eta^{(\prime)}}$ due to the smaller decay widths $\Gamma^{exp}_{J/\psi\rightarrow \rho\eta^{(\prime)}}$ and branching ratio $\mathcal{B}^{exp}(\rho\rightarrow e^{+}e^{-})$ as well as the larger full width $\Gamma_{\rho}$~\cite{ParticleDataGroup:2020ssz}. As pointed out in Ref~\cite{Intemann:1983yj}, there is still some open questions for the VMD model, such as the sign ambiguity in the generalized amplitude from the intermediate vector mesons ($\rho,\,\omega,\,\phi$) and the off-mass-shell effects of the coupling constants, and more discussions could be found in Refs.~\cite{Bauer:1977iq,Intemann:1983yj,Chernyak:1983ej,Landsberg:1986fd,Zhao:2010mm}.

\section{Results and discussions}
\label{sec:Results and discussions}

The $q^{2}$-dependent differential decay widths of $J/\psi\rightarrow\eta^{(\prime)}\ell^{+}\ell^{-}$ can be expressed as
\begin{eqnarray}\label{unnormalizedwid}
\frac{\mathrm{d}\Gamma(J/\psi\rightarrow\eta^{(\prime)}\ell^{+}\ell^{-})}{\mathrm{d}q^{2}}=\frac{1}{3}\frac{\alpha^{2}}{24\pi M^{3}}\frac{\mid f_{\psi \eta^{(\prime)}}(q^{2}) \mid^{2}}{q^{2}}\left(1+\frac{2m_{\ell}^{2}}{q^{2}}\right)
\left(1-\frac{4m_{\ell}^{2}}{q^{2}}\right)^{\frac{1}{2}}\lambda^{\frac{3}{2}}(M^{2}, \, m^{2}, \, q^{2}),
\end{eqnarray}
with $m_{\ell}$ the lepton mass. Here it should be noted that the TFFs $f_{\psi \eta^{(\prime)}}(q^{2})$ are taken the forms given in Eq.~\eqref{TFFs} in the numerical calculations without considering the VMD contributions; while in the numerical calculations including the hard, the soft and the VMD contributions, one can just take the replacement
\begin{eqnarray}
f_{\psi \eta^{(\prime)}}(q^{2})\to f_{\psi \eta^{(\prime)}}(q^{2})+\sum_{V=\rho,\omega,\phi}f^{V}_{\psi \eta^{(\prime)}}(q^{2}),
\end{eqnarray}
with the $f^{V}_{\psi \eta^{(\prime)}}(q^{2})$ given in Eq.~\eqref{TFFVMD}. In the following numerical calculations, the values of the involved meson masses, full widths, decay widths and decay constant are quoted from the PDG~\cite{ParticleDataGroup:2020ssz}. The QCD running coupling constant is adopted $\alpha_{s}(M/2)=0.34$, which is calculated through the two-loop renormalization group equation. As we have already mentioned, the theoretical uncertainties from $\eta^{(\prime)}$ DAs are negligible. So in our calculations, we choose the Model I of the meson DA in Table~1 of Refs.~\cite{He:2019mpy,Fan:2019sap}. For the value of the radial wave function at the origin $R_{\psi}(0)$, we adopt the result of the Cornell potential model~\cite{Eichten:1978tg,Eichten:1979ms,Eichten:1995ch}
\begin{eqnarray}
{\mid}R_{\psi}(0){\mid}^{2}=1.454 \, \mathrm{GeV}^{3}.
\end{eqnarray}

For $\eta-\eta^{\prime}$ system, in the quark-flavor basis, the effective decay constants $f_{\eta^{(\prime)}}^{q}$ are parameterized as~\cite{Akhoury:1987ed,Ball:1995zv,Feldmann:1998vh,Feldmann:1998sh,Feldmann:1999uf}
\begin{eqnarray}
f_{\eta}^{u(d)}&=&\frac{f_{q}}{\sqrt{2}}\cos\phi,\quad\quad   f_{\eta}^{s}=-f_{s}\sin\phi, \nonumber\\
f_{\eta^{\prime}}^{u(d)}&=&\frac{f_{q}}{\sqrt{2}}\sin\phi,\quad\quad   f_{\eta^{\prime}}^{s}=f_{s}\cos\phi.
\end{eqnarray}
Here the phenomenological parameters, i.e., the mixing angle $\phi$ and the decay constants $f_{q(s)}$, could be determined by different methods~\cite{Feldmann:1998vh,Gregory:2011sg,Michael:2013gka,Escribano:2013kba,Chen:2014yta,Escribano:2015nra,Escribano:2015yup,Urbach:2017rvx,He:2019mpy,Fan:2019sap,He:2020kin}. It is worth noting that this mixing scheme (i.e., Feldmann-Kroll-Stech scheme) arises as a special limit of the chiral Lagrangian, and more details and discussions could be found in Ref.~\cite{Mathieu:2010ss}. Usually, the value of mixing angle is in the range $\phi \sim (30^{\circ}-45^{\circ})$. In this work, we take the set of the parameter values~\cite{Escribano:2013kba}
\begin{eqnarray}
\phi=33.5^{\circ}\pm 0.9^{\circ},~~~~f_{q}=(1.09\pm 0.02)f_{\pi},~~~~f_{s}=(0.96\pm 0.04)f_{\pi},
\end{eqnarray}
which is compatible with the BABAR measurement~\cite{Aubert:2006cy} and consistent with the values obtained in other methods~\cite{Gregory:2011sg,Chen:2014yta}. In addition, our previous works~\cite{He:2019mpy,Fan:2019sap,He:2020kin} also indicated that a smaller value of the mixing angle $\phi$ ($\sim 34^{\circ}$) is favored in the radiative decays of charmonia with the framework of perturbative QCD.

With inputting all the parameters, we present the predictions of the branching ratios $\mathcal{B}(J/\psi\rightarrow\eta^{(\prime)}e^{+}e^{-})$ and their ratio $R_{J/\psi}^{e}$ in Table~\ref{tab:HS}. In the second column, we give the results without considering the VMD contributions; in the third column, we give the total results including the hard, the soft and the VMD contributions. Obviously, one can find that the VMD corrections are small in the branching ratios, especially for the $\eta^{\prime}$ channel. And the main reason is due to the very narrow peaks of resonances and the suppression of the kinematic factor (see Eq.~\eqref{unnormalizedwid}). In other words, this reminds us that an intuitive physical picture of the EM Dalitz decays of $J/\psi$ is mainly governed by the mesons' internal structure effects rather than the resonance effects. In addition, although the individual branching ratio is slightly smaller than the experiment data, their ratio agrees with its experiment data. This may imply that the higher order corrections related to the initial meson, such as the higher Fock-state contributions and the relativistic corrections, play an important role in these decay processes.

\begin{table}[!htbp]
  \caption{\label{tab:HS}The branching ratios $\mathcal{B}(J/\psi\rightarrow\eta^{(\prime)}e^{+}e^{-})$ and their ratio $R_{J/\psi}^{e}$.}
  \vspace{0.2cm}
  \centering
  \begin{tabular}{lccc}
  \hline\hline
  ~~&~~Without VMD~~~ &~~Total~~&~~Exp.~\cite{BESIII:2018qzg,Ablikim:2018bhf,ParticleDataGroup:2020ssz}~~~  \\
  \hline
 $\mathcal{B}(J/\psi\rightarrow\eta e^{+}e^{-})$~~&~~$0.96\times10^{-5}$~~&~~$1.05\times10^{-5}$~~&~~$(1.43\pm0.07)\times10^{-5}$ \\
 $\mathcal{B}(J/\psi\rightarrow\eta^{\prime}e^{+}e^{-})$~~&~~$4.00\times10^{-5}$
 ~~&~~$4.05\times10^{-5}$~~&~~$(6.59\pm0.18)\times10^{-5}$\\
 $R_{J/\psi}^{e}$                ~~&~~$24.1\%$~~&~~$25.9\%$ ~~&~~$(21.7\pm1.2)\%$  \\
  \hline\hline
  \end{tabular}
\end{table}

\begin{figure}[h]
  \begin{center}
  \includegraphics[width=0.45\textwidth]{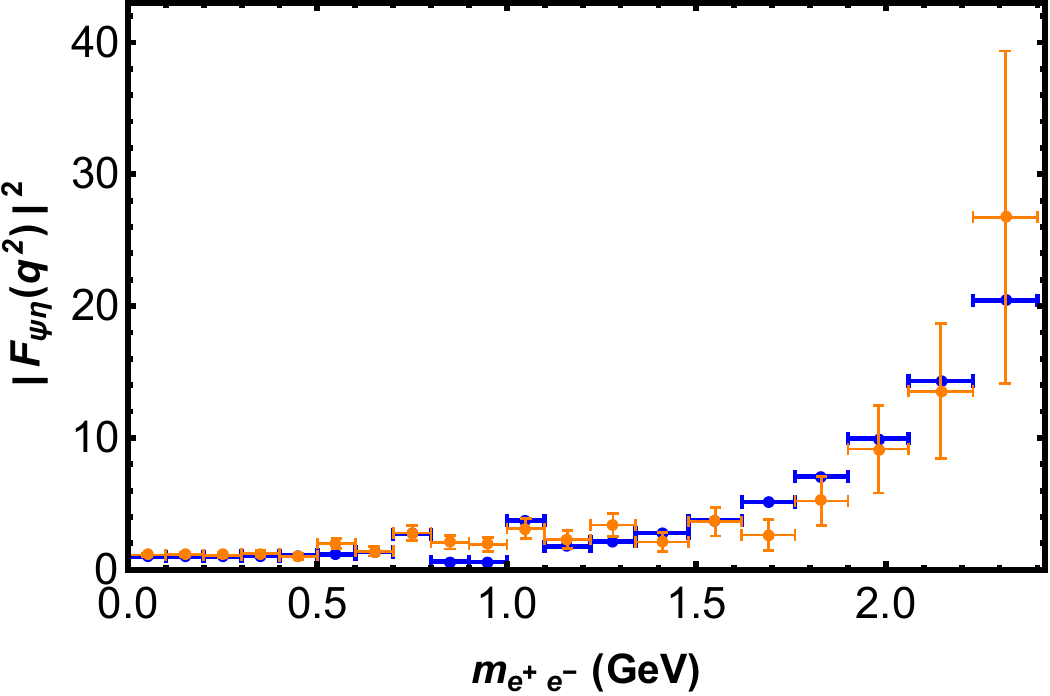}~~~~~~
  \includegraphics[width=0.45\textwidth]{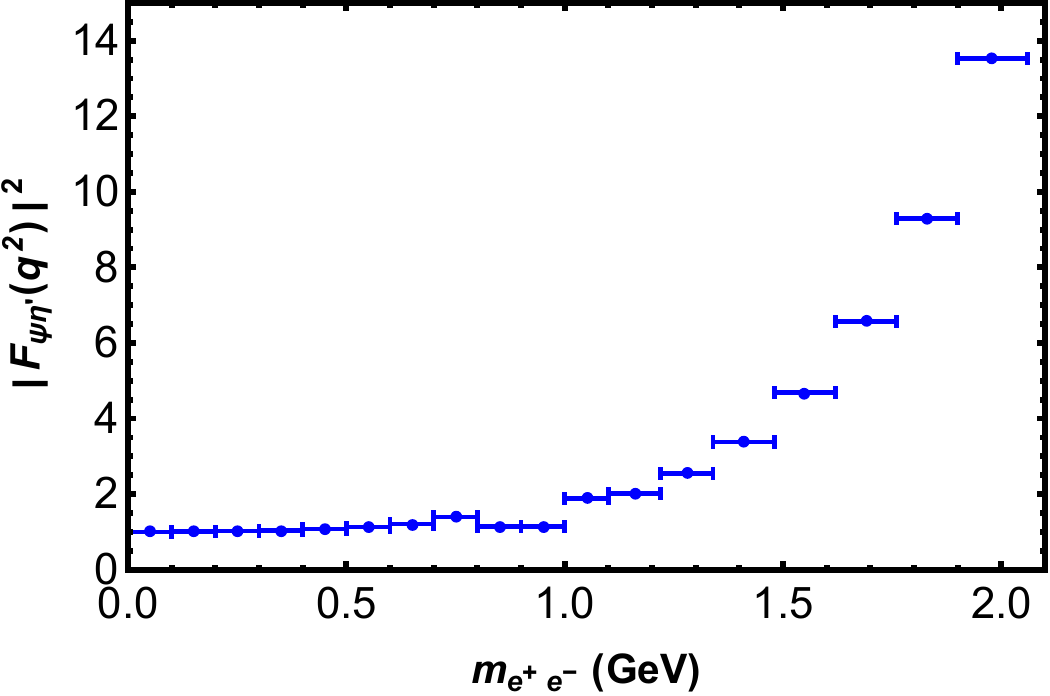}
  \end{center}
  \vskip -0.7cm
  \caption{The dependence of the modulus square of the normalized TFFs $|F_{\psi \eta^{(\prime)}}(q^{2})|^{2}$ on the dielectron invariant mass $m_{e^{+}e^{-}}$ (or, $q^{2}$). The blue dots with error bars are our results and the orange dots with error bars are experimental data~\cite{BESIII:2018qzg}.}\label{TFFHS}
\end{figure}

Besides the branching ratios, the study of the $q^{2}$-dependent TFFs is also important. It could provide more dynamical information on interactions between $J/\psi$ and the light mesons $\eta^{(\prime)}$, and offer a powerful probe of their internal structure. Given the uncertainties from the radial wave function at the origin $R_{\psi}(0)$, the coupling constant $\alpha_{s}(\mu)$ and the mixing angle $\phi$, one can relate the Dalitz decays $J/\psi\rightarrow\eta^{(\prime)}\ell^{+}\ell^{-}$ to the corresponding radiative decays $J/\psi\rightarrow\eta^{(\prime)}\gamma$ due to the similar dynamical properties, to lead to large cancellations of the uncertainties in the normalized TFFs $F_{\psi \eta^{(\prime)}}(q^{2})\equiv f_{\psi \eta^{(\prime)}}(q^{2})/f_{\psi \eta^{(\prime)}}(0)$. Even so, the $q^{2}$ dependence of the TFFs $f_{\psi \eta^{(\prime)}}(q^{2})$ is still retained in the normalized TFFs $F_{\psi \eta^{(\prime)}}(q^{2})$. In Fig.~\ref{TFFHS}, we show the $q^{2}$ dependence of the modulus square of the normalized TFFs $|F_{\psi \eta^{(\prime)}}(q^{2})|^{2}$ in the full kinematic region. In small $q^{2}$ region, we find that the $|F_{\psi \eta}(q^{2})|^{2}$ is quite steady, and it is in very nice agreement with the experimental measurement~\cite{BESIII:2018qzg}. This indicates that the hard mechanism gives a reliable description, in which both the quark-antiquark contributions and the gluonic contributions are included. While, in large $q^{2}$ region, the overlap of soft wave functions provides an intuitive physical picture, and its prediction is in accord with the recent BESIII measurement~\cite{BESIII:2018qzg}. For the intermediate region, the change tendency of the predicted $|F_{\psi \eta}(q^{2})|^{2}$ with VMD corrections is compatible with that of the experimental data~\cite{BESIII:2018qzg}. And it is worth noting that the small peaks correspond to the contributions from the intermediate vector mesons. In addition, the peaks in the $\eta^{\prime}$ channel are smaller than those in the $\eta$ channel mainly due to the coupling constants $g_{\psi V\eta^{\prime}}<g_{\psi V\eta}$, which leads to the suppression of the VMD contributions in the $\eta^{\prime}$ channel. The small discrepancies in some certain bins maybe due to the sign ambiguity of the amplitude from VMD model~\cite{Intemann:1983yj} or/and more resonance effects (such as the excited vector mesons or other exotic hadronic states~\cite{BESIII:2019dme}). As discussed before, although the VMD contributions are negligibly small in the branching ratios purely due to a suppression of the kinematic factor, they play a limited but significant role in the TFFs, and the physical picture in intermediate region is certainly worth further investigations.

\begin{figure}[h]
  \begin{center}
  \includegraphics[width=0.45\textwidth]{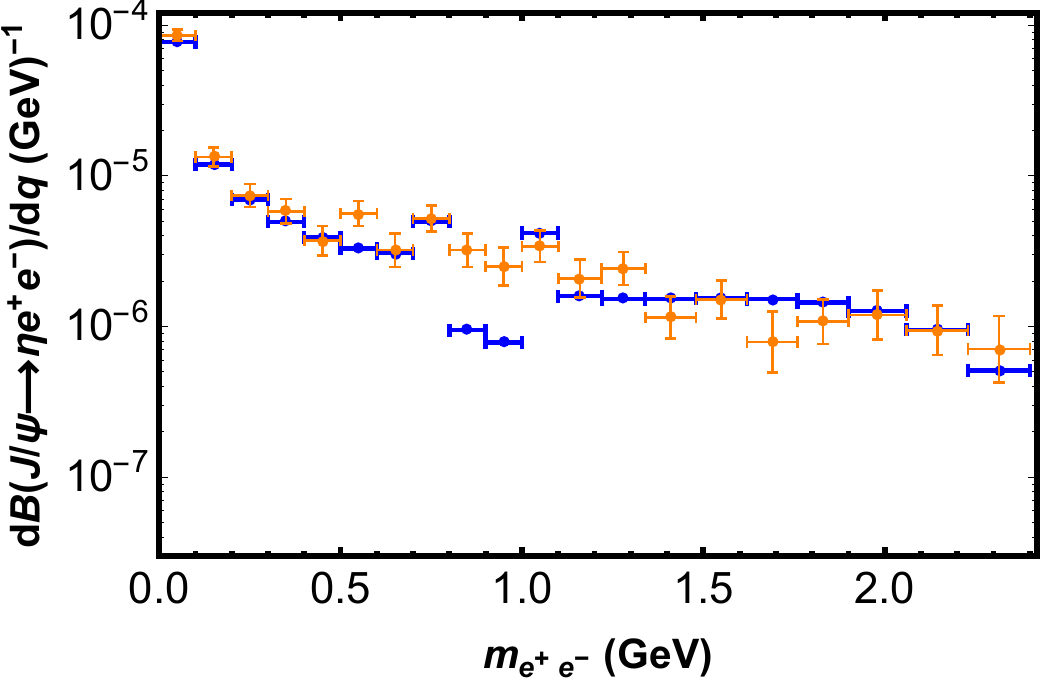}~~~~~~
  \includegraphics[width=0.45\textwidth]{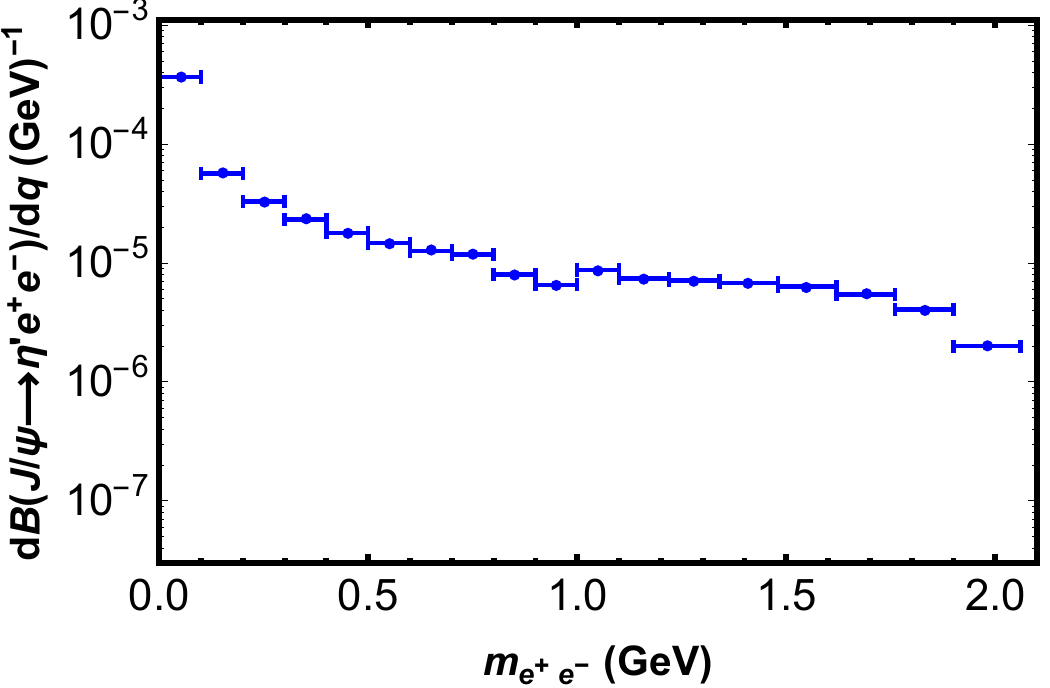}
  \end{center}
  \vskip -0.7cm
  \caption{The differential branching ratios of $J/\psi\rightarrow\eta^{(\prime)} e^{+}e^{-}$ in the full kinematic region. The blue dots with error bars are our results and the orange dots with error bars are experimental data~\cite{BESIII:2018qzg}.}\label{dbra}
\end{figure}

By employing the normalized TFFs $F_{\psi \eta^{(\prime)}}(q^{2})$, the differential branching ratios of $J/\psi\rightarrow\eta^{(\prime)} e^{+}e^{-}$ are given in Fig.~\ref{dbra}. Similarly to the situation in our discussion about the normalized TFFs, an excellent agreement between calculated and experimental data is obtained in small and large $q^{2}$ regions, and the change tendency of the predicted differential branching ratio $\mathrm{d}\mathcal{B}(J/\psi\rightarrow\eta e^{+}e^{-})/\mathrm{d}q$ is compatible with the experimental data in the intermediate region, where the small peaks correspond to the contributions from the intermediate vector mesons. As shown in Fig.~\ref{dbra}, one can find that the contributions of perturbative QCD from the large recoil momentum region (i.e., small $q^{2}$) are much larger than the ones from other region. Specifically, the perturbative QCD contributions are an order of magnitude larger than the VMD and the soft contributions. The main reason is that the differential branching ratio is proportional to $|\mathbf{p}_{\eta^{(\prime)}}|^{3}/q$, i.e.,
\begin{eqnarray}
\frac{\mathrm{d}\mathcal{B}(J/\psi\rightarrow\eta^{(\prime)} e^{+}e^{-})}{\mathrm{d}q}\propto \frac{|\mathbf{p}_{\eta^{(\prime)}}|^{3}}{q},
\end{eqnarray}
where $|\mathbf{p}_{\eta^{(\prime)}}|=\lambda^{\frac{1}{2}}(M^{2}, \, m^{2}, \, q^{2})/(2M)$ (easily, $q$ is a monotonically decreasing function of $|\mathbf{p}_{\eta^{(\prime)}}|$) is the recoil momentum of $\eta^{(\prime)}$ in the rest frame of $J/\psi$, and just this kinematic factor leads to a strong suppression as $|\mathbf{p}_{\eta^{(\prime)}}|$ decrease. Performing the corresponding integration, we present the results of the branching ratios $\mathcal{B}(J/\psi\rightarrow\eta^{(\prime)}e^{+}e^{-})$ and their ratio $R_{J/\psi}^{e}$ in Table~\ref{tab:normalizedTFF}. It is found that not only the ratio $R_{J/\psi}^{e}$ but also the individual branching ratios $\mathcal{B}(J/\psi\rightarrow\eta e^{+}e^{-})$ and $\mathcal{B}(J/\psi\rightarrow\eta^{\prime}e^{+}e^{-})$ are in nice agreement with their experiment data. Comparing the results listed in Table~\ref{tab:HS} with those listed in Table~\ref{tab:normalizedTFF}, we find that both the branching ratios $\mathcal{B}(J/\psi\rightarrow\eta e^{+}e^{-})$ and $\mathcal{B}(J/\psi\rightarrow\eta^{\prime}e^{+}e^{-})$ are enhanced considerably. This indicates that the dynamical effects from the initial meson $J/\psi$ may play a key role in both its EM Dalitz decays and the corresponding radiative decays. And similar effects have been studied in the radiative decays of $h_{c}$~\cite{Fan:2019sap,He:2020kin}.

\begin{table}[!htbp]
  \caption{\label{tab:normalizedTFF}The branching ratios $\mathcal{B}(J/\psi\rightarrow\eta^{(\prime)}e^{+}e^{-})$ and their ratio $R_{J/\psi}^{e}$ with the normalized TFFs.}
  \vspace{0.2cm}
  \centering
  \begin{tabular}{lcc}
  \hline\hline
  ~~&~~Theo. ~~&~~Exp. ~\cite{BESIII:2018qzg,Ablikim:2018bhf,ParticleDataGroup:2020ssz}~~~  \\
  \hline
 $\mathcal{B}(J/\psi\rightarrow\eta e^{+}e^{-})$~~&~~$1.38\times10^{-5}$~~&~~$(1.43\pm0.07)\times10^{-5}$ \\
 $\mathcal{B}(J/\psi\rightarrow\eta^{\prime}e^{+}e^{-})$~~&~~$6.06\times10^{-5}$~~&~~$(6.59\pm0.18)\times10^{-5}$\\
 $R_{J/\psi}^{e}$                ~~&~~$22.7\%$ ~~&~~$(21.7\pm1.2)\%$  \\
  \hline\hline
  \end{tabular}
\end{table}

Furthermore, we give our predictions of the branching ratios $\mathcal{B}(J/\psi\rightarrow\eta\mu^{+}\mu^{-})$, $\mathcal{B}(J/\psi\rightarrow\eta^{\prime}\mu^{+}\mu^{-})$ and their ratio $R_{J/\psi}^{\mu}$ employing the normalized TFF $F_{\psi \eta^{(\prime)}}(q^{2})$:
\begin{eqnarray}
\mathcal{B}(J/\psi\rightarrow\eta\mu^{+}\mu^{-})=4.61\times10^{-6},\quad
\mathcal{B}(J/\psi\rightarrow\eta^{\prime}\mu^{+}\mu^{-})=1.72\times10^{-5},\quad
R_{J/\psi}^{\mu}=26.7\%.
\end{eqnarray}
Clearly, the branching ratios of the $\mu^{+}\mu^{-}$ channels are much smaller than those of the $e^{+}e^{-}$ channels, and the main reason is due to the shrinking of phase space. Future experimental measurement is expected to provide tests for these predictions.

Besides, the EM Dalitz decays $J/\psi\rightarrow \pi^{0} \ell^{+}\ell^{-}$ have been studied in the experimental aspect~\cite{BESIII:2014dax} and the theoretical aspect (such as the effective Lagrangian approach~\cite{Chen:2014yta} and the dispersion theory~\cite{Kubis:2014gka}). Similar to the situation in the radiative decay $J/\psi\rightarrow \pi^{0}\gamma$~\cite{He:2019mpy}, we find that the one-loop QCD contributions to the Dalitz decays $J/\psi\rightarrow \pi^{0} \ell^{+}\ell^{-}$ almost vanish as a consequence of the antisymmetrical flavor wave function of $\pi^{0}$ and the QED contributions are an order of magnitude smaller than the experimental data. As pointed out in Ref.~\cite{Chernyak:1983ej}, the processes $J/\psi \rightarrow \pi^{0} \ell^{+}\ell^{-}$, as well as $J/\psi\rightarrow \pi^{0}\gamma$, may be dominated by the two-loop QCD contributions which are beyond the scope of this paper.

\section{Summary}
\label{sec:conclusion}
In this work, we present a QCD analysis of the EM Dalitz decays $J/\psi\rightarrow\eta^{(\prime)}\ell^{+}\ell^{-}$, and propose a dynamical description in the full kinematic region. In the large recoil momentum region, these processes are studied in detail with the perturbative QCD approach, in which the internal structure effects of $J/\psi$ are absorbed into its bound state wave function and the light mesons $\eta^{(\prime)}$ are described by their light-cone DAs due to the large recoil momentum. In the small recoil momentum region, the picture of the soft wave function overlap is adopted to describe these transition processes. In the intermediate region, the resonance contributions are estimated by the VMD model. Based on this intuitive physical picture, the contributions from hard mechanism, soft mechanism and VMD model in these processes are explored for the first time. We find that the branching ratios $\mathcal{B}(J/\psi\rightarrow\eta^{(\prime)}\ell^{+}\ell^{-})$ are dominated by the perturbative QCD contributions from the hard mechanism, while the contributions from the soft mechanism and the VMD model are strongly suppressed by a kinematic factor. By relating to their radiative decay processes, the branching ratios $\mathcal{B}(J/\psi\rightarrow\eta^{(\prime)}e^{+}e^{-})$ and the ratio $R_{J/\psi}^{e}$ are in excellent agreement with the experimental measurements, and then the predictions for the $\mu^{+}\mu^{-}$ channels are given.

Furthermore, the $q^{2}$-dependent TFFs and differential branching ratios are analysed in detail. It is found that the TFFs with small $q^{2}$ are insensitive to the light quark masses and the shapes of the $\eta^{(\prime)}$ DAs, which is in line with the conclusion given in our previous works~\cite{He:2019mpy,Fan:2019sap,He:2020kin}. As shown in the numerical analysis, our predictions of the TFFs and the differential branching ratios are in good agreement with the experimental data in small and large $q^{2}$ regions, and the change tendency of the predictions in the intermediate region is compatible with the experimental data. And this further indicates that the dynamical picture adopted by us is reasonable. In addition, this dynamical description is quite general, it can be widely used to dynamically study the similar decays of charmonia.

Lastly, it should be pointed out that there exist the radiative corrections in the case of electron pair production. However, a full analysis of radiative corrections seems highly impractical since it would require a two-loop calculation. Of course, the radiative corrections are negligibly small in the case of muon pair production~\cite{Landsberg:1986fd}. Anyhow, these EM Dalitz decay processes not only provide rich dynamical information about the charmonium physics, but also could be used for exploring the nature of exotic hadrons~\cite{BESIII:2021xoh} and testing the lepton-flavor universality (see~\cite{Albrecht:2021tul} and references therein for more details). And it deserves further experimental and theoretical studies.

\section*{Acknowledgements}
This work is supported by the National Natural Science Foundation of China under Grant Nos.~11675061, 11775092 and 11435003.

\newpage

\bibliographystyle{JHEP}
\bibliography{hejk}

\end{document}